\documentclass[%
reprint,
superscriptaddress,
showpacs,
showkeys,
 amsmath,amssymb,
aps,
pra,
longbibliography,
floatfix
]{revtex4-1}

\usepackage{graphicx}% Include figure files
\usepackage{dcolumn}% Align table columns on decimal point

\usepackage{bm}% bold math
\usepackage[english]{babel}
\usepackage[colorinlistoftodos]{todonotes}

\usepackage[utf8]{inputenc}
\usepackage[T1]{fontenc}
\usepackage{mathtools}
\DeclarePairedDelimiter{\ceil}{\lceil}{\rceil}

\usepackage{natbib}
\usepackage{letltxmacro}

\LetLtxMacro{\ORIGselectlanguage}{\selectlanguage}
\makeatletter
\DeclareRobustCommand{\selectlanguage}[1]{%
  \@ifundefined{alias@\string#1}
    {\ORIGselectlanguage{#1}}
    {\begingroup\edef\x{\endgroup
       \noexpand\ORIGselectlanguage{\@nameuse{alias@#1}}}\x}%
}
\newcommand{\definelanguagealias}[2]{%
  \@namedef{alias@#1}{#2}%
}
\makeatother

\definelanguagealias{en}{english}

\newcommand{\rref}[1]{Eq.\ (\ref{#1})}
\newcommand{\rrefsa}[1]{Eqs.\ (\ref{#1})}
\newcommand{\rrefsb}[1]{(\ref{#1})}

\usepackage{color}
\usepackage[unicode]{hyperref}
\hypersetup{
   unicode=true,          % non-Latin characters in Acrobat??s bookmarks
   plainpages=false,
   colorlinks=true,       % false: boxed links; true: colored links
   citecolor=blue,        % color of links to bibliography
}

\usepackage{url}

\begin{document}

\title{Are smooth pseudopotentials a good choice for representing
short-range interactions?
%Simple expressions to determine the minimal number of basis functions for the exponential convergence  of Gauss pseudopotential
%\todo{In the title we need harmonic trap.}
}

\date{\today}% It is always \today, today,
             %  but any date may be explicitly specified

\author{P\'eter Jeszenszki}
 \email{jeszenszki.peter@gmail.com}
 \affiliation{Dodd-Walls Centre for Photonics and Quantum Technology, PO Box 56, Dunedin 9056, New Zealand}
\affiliation{%
New Zealand Institute for Advanced Study, and Centre for Theoretical Chemistry and Physics, Massey University, Private Bag 102904 North Shore, Auckland 0745, New Zealand
}%
\affiliation{
 Max Planck Institute for Solid State Research, Heisenbergstra{\ss}e 1,
70569 Stuttgart, Germany
}%

%\collaboration{MUSO Collaboration}%\noaffiliation

%\author{Tal Levy}
%\email{j.l.greystoke@gmail.com }
%\affiliation{
% Max Planck Institute for Solid State Research, Heisenbergstra{\ss}e 1,
%70569 Stuttgart, Germany
%}%
\author{Ali Alavi}
\email{A.Alavi@fkf.mpg.de}
\affiliation{
 Max Planck Institute for Solid State Research, Heisenbergstra{\ss}e 1,
70569 Stuttgart, Germany
}%
 \affiliation{Department of Chemistry, University of Cambridge,
Lensfield Road, Cambridge, CB2 1EW, United Kingdom}

\author{Joachim Brand}%
 \email{J.Brand@massey.ac.nz}
 \affiliation{Dodd-Walls Centre for Photonics and Quantum Technology, PO Box 56, Dunedin 9056, New Zealand}
\affiliation{%
New Zealand Institute for Advanced Study, and Centre for Theoretical Chemistry and Physics, Massey University, Private Bag 102904 North Shore, Auckland 0745, New Zealand
}%
\affiliation{
 Max Planck Institute for Solid State Research, Heisenbergstra{\ss}e 1,
70569 Stuttgart, Germany
}%

\begin{abstract}
When seeking a numerical representation of a quantum-mechanical multiparticle problem it is tempting to replace a singular short-range interaction by a smooth finite-range pseudopotential. Finite basis set expansions, e.g., in Fock space, are then guaranteed to converge exponentially. The need to faithfully represent the artificial length scale of the pseudopotential, however, places a costly burden on the basis set. Here we discuss scaling relations for the required size of the basis set and demonstrate the basis set convergence on the example of a two-dimensional system of few fermions with short-range $s$-wave interactions in a harmonic trapping potential. In particular we show that the number of harmonic-oscillator basis functions needed to reach a regime of exponential convergence for a Gaussian pseudopotential scales with the fourth power of the pseudopotential length scale, which can be improved to quadratic scaling when the basis functions are rescaled appropriately.
Numerical examples for three fermions with up to a few hundred single-particle basis functions are presented and implications for the feasibility of accurate numerical multiparticle simulations of interacting ultracold-atom systems are discussed.

%original version
% The Gauss interaction potential is a widely used finite-range pseudopotential for ultracold atoms. Applying an exact diagonalization method, the wave function and the energy is expected to converge exponentially fast 
% as the potential is smooth. Considering a small number of basis functions the finite-range potential indistinguishable from the Dirac delta potential, which is pathological in two and three dimensions. We derive a simple expression for a minimal number
% of single-particle harmonic oscillator basis functions, which are necessary to resolve the smooth structure of the Gaussian potential leading to exponentially fast convergence. The single-particle basis is also optimized by scaling the unit length of the basis.
% We show that the unscaled and the scaled harmonic oscillator basis functions have 
% quantic and quadratic dependence on the characteristic length scale of the Gaussian potential. 
 %These findings are presented in an example of three fermions in two-dimensional harmonic oscillator.   

\end{abstract}

%\pacs{Valid PACS appear here}% PACS, the Physics and Astronomy
                             % Classification Scheme.
%Use showkeys class option if keyword
                              %display desired
\maketitle

%\tableofcontents

\section{Introduction}

There is increasing interest in the theoretical description of multiparticle systems of interacting ultracold atoms  thanks to the recent progress in experimental realizations
\cite{cheuk_observation_2016,feld_observation_2011,zurn_pairing_2013,mazurenko_cold-atom_2017,hueck_two-dimensional_2018}.
In particular we may expect exciting developments in microtraps \cite{serwane_deterministic_2011, zurn_fermionization_2012,kaufman_two-particle_2014} with tens of particles where accessing strongly correlated regimes of quantum-Hall-like physics seems feasible \cite{Popp2004,viefers_quantum_2008, regnault_quantum_2004,moller_pairing_2009}. %bencheikh_current_2014}.

The theoretical description of atom-atom interactions is significantly simplified  at ultracold temperatures where details of the  interaction potentials can be neglected in favor of a single constant, the $s$-wave scattering length $a_s$, to define a physical model with contact interactions \cite{pethick_bose-einstein_2002}. Despite these simplifications, the complexity of many-particle quantum mechanics still makes it a very difficult problem to solve, where exact solutions are only available in special cases in one spatial dimension \cite{lieb_exact_1963,yang_exact_1967,gaudin_systeme_1967,volosniev_strongly_2014,deuretzbacher_exact_2008} or for up to three particles in a harmonic trap \cite{Busch1998,Liu2010,liu_exact_2010}.

A straightforward and generalist approach to representing the many-body problem for computational treatment is to introduce a discrete and necessarily finite basis of smooth single-particle wave functions from which a finite but still potentially very large Fock-space is constructed to represent the many-body Hamiltonian as a matrix. Finding eigenstates and eigenvalues of the full matrix is known as exact diagonalization or full configuration-interaction \cite{Deuretzbacher2007,rontani_cold_2009,garcia-march_sharp_2013,dehkharghani_quantum_2015,mujal_quantum_2017}, but many different approximation schemes have also been followed \cite{marcin_plodzien_numerically_2018}. 
In particular, standard approaches of {\it ab initio} quantum chemistry or nuclear physics like the coupled-cluster \cite{grining_many_2015} or multi configurational self-consistent field theory \cite{Meyer2009} all can be formulated in this language as they rely on an underlying single-particle basis. Also Monte Carlo (or other) approaches that rely on a lattice discretization of continuous space fall into the same category  \cite{Carlson2011}, as the underlying single-particle space can be represented as a discrete set of plane waves.

%In homogeneous case or in a harmonic trap the number of variables can be reduced by transforming out the centre-of-mass part of the wave function, which can be trivially solved. The relative motion part  can be 
%elegantly solved for three particles in one dimension by simplifying with geometrical point groups the permutational symmetry of the wave function to the particle exchange  \cite{harshman_symmetries_2012,garcia-march_distinguishability_2014,blume_few-body_2012}. In this case, the basis functions depend on the coordinate of the relative motion providing a faster converging expansion. However, this approach becomes more difficult by increasing the particle number and the dimension of the spatial space \cite{von_stecher_correlated_2009,blume_few-body_2012,yin_trapped_2015}. Alternatively, a hyperspherical method is applied

One of the complications in the numerical treatment of contact interactions with basis set expansions stems from the nonanalytic behavior of the wave function at the point of particle coalescence.
At this point, the appropriate Bethe-Peierls boundary conditions demand a cusp in one spatial dimension, i.e.~a point of nondifferentiability \cite{lieb_exact_1963}, 
in two dimensions  a logarithmic divergence, and  in three dimension a $1/r$ divergence of the wave function
%and the appropriate Bethe-Peierls boundary conditions demand a logarithmic divergence in two dimensions and a $1/r$ divergence of the wave function in three dimension 
\cite{bethe_quantum_1935,olshanii_short-distance_2003,werner_general_2012}. While in one dimension a Dirac $\delta$ function pseudopotential provides a well-defined model for contact interactions, the convergence of basis set expansions is algebraic and painfully slow  \cite{grining_many_2015,jeszenszki_accelerating_2018}.  Basis set expansions in two and three dimensions diverge for bare contact interaction \cite{esry_validity_1999,doganov_two_2013,rontani_renormalization_2017} and  basis-set-dependent renormalization procedures have to be used in order to obtain convergent and correct results \cite{castin_simple_2004,ernst_simulating_2011,rontani_renormalization_2017, luo_harmonically_2016}. In the best case, renormalized contact interactions will lead to algebraic convergence in the size of the finite single-particle basis set \cite{Carlson2011,werner_general_2012,jeszenszki_accelerating_2018}. Some of us have recently described a transcorrelated method where the singular nature of the contact interaction is reduced by means of a similarity transformation of the Hamiltonian \cite{jeszenszki_accelerating_2018} (see also Ref.~\cite{Rotureau2013} for a related idea). This promising approach results in an improved power-law scaling but the convergence still remains algebraic.

Here, we consider a different approach where the contact interaction is replaced by a smooth finite-range pseudopotential. This has the advantage that basis set expansions will converge exponentially for appropriately chosen single-particle basis functions \cite{mercier_introduction_2014}. Specifically we consider what the requirements are for the basis set to reach the regime of exponential convergence and whether this approach is feasible for multiparticle simulations. Examples of finite-range pseudopotentials used in the literature are the
Troullier-Martins  \cite{bugnion_high-fidelity_2014,whitehead_pseudopotential_2016},
Pöschl-Teller \cite{forbes_resonantly_2011,galea_diffusion_2016}, and Gaussian potentials \cite{christensson_effective-interaction_2009,von_stecher_energetics_2008,von_stecher_bec-bcs_2007,yan_abnormal_2014,yin_harmonically_2014,doganov_two_2013,imran_exact_2015,klaiman_breaking_2014, beinke_many-body_2015,bolsinger_beyond_2017,bolsinger_ultracold_2017,mujal_quantum_2017,mujal_fermionic_2018}, in addition to the square well popular in diffusion Monte Carlo simulations\cite{von_stecher_energetics_2008}. 

When using finite range pseudopotentials to represent short-range interactions, an interpolation in the width of the pseudopotential should be done  to the zero-range limit \cite{blume_few-body_2012}. In order to approach this limit, the length scale of the pseudopotential should be significantly smaller than other physically relevant length scales of the problem, in particular the mean particle separation and length scales imposed by external potentials. In order to reach a regime where the basis set expansion converges exponentially, however, the basis set needs to resolve the smallest length scale of the pseudopotential. %Otherwise, if the basis set cannot resolve the structure of the pseudopotential, the 
At the same time, the large length scales of the problem, i.e., the (Thomas-Fermi) size of the cold atom cloud, or the size of the container, also have to be represented by the basis set. This hierarchy of length scales, typically spanning at least one but possibly several orders of magnitude, presents a challenge for accurate numerical simulations. While the size of the single-particle basis (quantified by the number of single-particle functions $M$) is determined by this hierarchy of length scales, the size of the full many-body problem also depends strongly on the number of particles $N$. Specifically for spinless bosons, the size of the relevant part of Fock space is $\binom{N+M-1}{M}$, whereas for spin-$\frac{1}{2}$ fermions the total dimension is $\binom{M}{N_\uparrow}\binom{M}{N_\downarrow}$, where $N_\uparrow$ and $N_\downarrow$ are the numbers of up- and down-spin particles, respectively.

In this work, we specifically consider ultracold fermionic atoms in a harmonic oscillator trapping potential where the potential in one of the three trapping directions is so tight that the problem can be considered two-dimensional. We furthermore choose a Gaussian pseudopotential to model attractive $s$-wave interactions between spin-up and spin-down particles \cite{jeszenszki_s_2018}. For the underlying single-particle basis we consider two cases: (1) a basis that is defined by the single-particle eigenstates of the isotropic two-dimensional harmonic trapping potential, and (2) the same set of basis functions with scaled spatial coordinates by a scaling factor $\gamma$. Using the known properties of the harmonic oscillator eigenfunctions we show that the basis set size $M$ required to resolve the chosen length scale of the pseudopotential $l_\mathrm{res}$ scales as $(l/l_\mathrm{res})^4$ where $l = \sqrt{\hbar/m\omega}$ is the harmonic oscillator length scale of the trapping potential for case (1). 

Allowing the basis functions to be scaled by $\gamma$ as in case (2) leads to an improved scaling of $(l/l_\mathrm{res})^2$ while still faithfully resolving the small length scale $l_\mathrm{res}$ and a fixed large length scale that is determined by the particle-number and interaction strength. We provide estimates for these length scales and the required scaling parameters. Numerical examples show the convergence of the ground-state energy for three fermions obtained by exact diagonalization with single-particle basis sets of up to 231 Fock-Darwin orbitals. In order to compute the matrix elements of the Gaussian interaction potential with the  Fock-Darwin basis of this size, a careful algorithm based on recursion formulas had to be developed in order to avoid an excessive accumulation of round-off errors. This algorithm is described in Appendix \ref{appendixint}.

This paper is organized as follows:
After defining the Hamiltonian in Sec.~\ref{sec:Hamiltonian} and introducing the methodology in Sec.~\ref{sec:numerics}, we discuss the main results of the paper in Sec.~\ref{sec:resol}.  
Examples for the numerical convergence with a harmonic oscillator basis are presented in Sec.~\ref{sec:unscaled} before deriving analytical formulas for the required minimum basis set size in  Sec.~\ref{sec:length} for the unscaled and in Sec.~\ref{sec:scaled} for a scaled harmonic oscillator basis. The required scaling factor is estimated in Sec.~\ref{sec:prefactor} where also numerical results for the scaled basis are presented.
%The convergence properties are considered by analytical arguments and  with numerical examples for the  unscaled harmonic-oscillator basis in Sec.~\ref{sec:conv} and for the scaled basis in Sec.~\ref{sec:scaled} before estimating the scaling factor in Sec.~\ref{sec:prefactor}. 
Implications of our findings for the feasibility of accurate computations of larger multiparticle problems are discussed in Sec.~\ref{sec:concl}. Two appendixes define the Fock-Darwin orbital basis used (Appendix \ref{app:basis}) and detail the explicit formulas and the algorithm used to compute the matrix elements  (Appendix \ref{appendixint}).

\section{Hamiltonian}\label{sec:Hamiltonian}

We consider ultracold fermions in a two-dimensional harmonic trap,
%which can be described by the following Hamiltonian: 
\begin{eqnarray} \label{eq:Hfull}
H &=& H_{\mathrm{osc}} + \sum_i^{N_\uparrow} \sum_j^{N_\downarrow} V({\bf r}_{i\uparrow},{\bf r}_{j\downarrow}) \ , \\
H_{\mathrm{osc}} &=& \sum_{\sigma}\sum_i^{N_\sigma}  \left(  -\frac{\hbar^2}{2m} \nabla^2_{i \sigma}  \ + \ \frac{m \omega^2}{2} r_{i \sigma}^2  \right) \ , \label{Hoscdef} 
\end{eqnarray}
where $m$ is the mass of the fermions, $\omega$ is the harmonic 
oscillator strength, $r_{i\sigma}$ is the position of the 
$i$th particle with the spin $\sigma$, and $V({\bf r}_{i\uparrow},{\bf r}_{j\downarrow})$ is the interaction potential between the fermions. 
Dividing the operator $H_{\mathrm{osc}}$ with $\hbar \omega$,  \rref{Hoscdef} takes the form 
\begin{eqnarray*}
\frac{H_{\mathrm{osc}}}{\hbar \omega} &=& \sum_{\sigma}\sum_i^{N_\sigma}  \left(  -\frac{l^2}{2} \nabla^2_{i \sigma}  \ + \ \frac{1}{2l^2} r_{i \sigma}^2  \right) \ ,
\end{eqnarray*}
where $l=\sqrt{\frac{\hbar}{m \omega}}$ is the harmonic oscillator length scale of the trapping potential.
%introduced, which gives the characteristic length scale of the system. 
The interaction between the particles is described with a Gaussian pseudopotential
\begin{eqnarray}
V({\bf r}_{i\uparrow},{\bf r}_{j\downarrow}) &=& 
-\frac{V_0}{R^2} \, e^{-\frac{\left({\bf r}_{i \uparrow}-{ \bf r}_{j \downarrow} \right)^2}{R^2}} \ . 
\end{eqnarray}
The parameters $V_0$ and $R$ control the strength and width of the interaction potential, respectively. 
These parameters can be converted to the $s$-wave scattering length using a simple approximate formula  \cite{jeszenszki_s_2018},
%with approximate formulas or by direct numerical computation as is discussed in detail in Ref.~\cite{jeszenszki_s_2018}.
\begin{eqnarray}
\frac{a_s}{R} &=& \sqrt{2}  \, \exp \left( -\frac{3\gamma_E}{2}+\frac{8\hbar^2}{V_0 m } + 
\sum_i^n  \alpha_i \frac{ V_0}{V_0-W_i}\right) \ ,
\label{swavelengthexpr}
\end{eqnarray}
where $a_s$ is the $s$-wave scattering length in two dimensions, $\gamma_E$ is the Euler-Mascheroni constant with the approximate value  of $\gamma_\mathrm{E} \approx 0.577216$, and $\alpha_i$ and $W_i$ are parameters fitted to direct numerical calculations \footnote{ Equation
\rrefsb{swavelengthexpr} slightly differs from Eq. (36) in Ref. \cite{jeszenszki_s_2018} as they are written in different units. The two equations can be transformed to each other by considering the relation between the relative mass and the mass of the particle, $\mu=m/2$}. 
Accurate numerical values of the parameters are given in Table III. in Ref. \cite{jeszenszki_s_2018} for $i \le 4$.
Alternatively, more complicated numerical approaches can be applied \cite{verhaar_scattering_1985,jeszenszki_s_2018}. 
%The connection between the parameters 
%and the actual physical potential can be made by determining the $s$-wave scattering length \cite{jeszenszki_s_2018,christensson_effective-interaction_2009}.

%{\color{red} Maybe something about large difference between the two length scales \cite{cosme_center--mass_2016}? I can separate the center of mass for the relative motion, where it can be explicitly shown that even if the Gauss potential is large compare to the harmonic oscillator the center-of-mass independent from that. Hence, I have to treat both of the length scales in the strongly attractive limit.}

\section{Basis-set expansion}\label{sec:numerics}

For our numerical approach, we compute the ground-state energy of a multiple-fermion system following the exact diagonalization approach. Starting from a finite single-particle basis of size $M$ (i.e., with $M$ spin orbitals), the multiparticle wave function is expanded as a linear combination
\begin{align}
| \Psi \rangle = \sum_{\underbar{n}} C_{\underbar{n}} |\Phi_{\underbar{n}} \rangle , \label{exactdiag}
\end{align}
of states in the Fock basis
\begin{align} \label{eq:FockStates}
 |\Phi_{\underbar{n}} \rangle = \displaystyle\prod_{i=1}^M (\hat{c}_i^\dag)^{n_i} |\mathrm{vac}\rangle ,
\end{align}
where $\underbar{n} = (n_1,\ldots, n_i,\ldots, n_M)$ and $n_i$ is the occupation number of the $n$th single-particle basis function (spinr orbital). The fermionic Fock states $ |\Phi_{\underbar{n}} \rangle$ (often referred to as Slater determinants) are constructed from the complete set of index vectors $\underbar{n}$ for fixed particle number $N$, with
\begin{align}
N = \sum_{i=1}^M n_i .
\end{align}
The exact diagonalization approach (also referred to as full configuration interaction) refers to considering the multiparticle Hamiltonian \eqref{eq:Hfull} with the chosen particle-number content projected onto the basis of states \eqref{eq:FockStates} as a matrix and finding its eigenvalues and eigenvectors.

%We apply the exact diagonalization approach.
%The wave function is expanded in the basis of Fock-space vectors,
%\begin{eqnarray}
%| \Psi \rangle = \sum_i c_i  | \Phi_i \rangle \ , \label{exactdiag}
%\end{eqnarray}
%where $| \Phi_i \rangle$ is a Fock-space vector and  $c_i$ is the corresponding coefficient. 
%For fermions, the Fock-space basis vector is frequently referred to as Slater determinants as it is given by an anti-symmetric product of single-particle functions. After choosing a finite set of these single-particle functions all the possible determinants are generated with a definite particle number in expansion \rref{exactdiag}. Then the coefficients $c_i$ can be obtained by evaluating and the diagonalizing the Hamiltonian in this many-body basis.  

In this paper we use a single-particle basis constructed from the spinful eigenstates of the two-dimensional harmonic oscillator \eqref{Hoscdef}.
% eigenstates augmented with simple spin-up or spin-down spinors. 
%functions are applied as a single-particle basis. Its 
The explicit form of the basis functions used in the numerical procedure is presented in Appendix \ref{app:basis}. Even though the one-body and two-body integrals needed for the relevant matrix elements can be expressed analytically (see, e.g.,  Ref. \cite{mujal_quantum_2017}), obtaining accurate numerical values is challenging due to the proliferation of rounding errors during floating-point arithmetic. We have therefore developed an iterative algorithm for the evaluation of the two-body integrals, which alleviates this problem. The details are presented in Appendix  \ref{appendixint}.

% and the determination of the corresponding matrix elements are discussed in detail
%% in Ref. \cite{mujal_quantum_2017} and 
%in Appendices \ref{app:basis} and \ref{appendixint}.

For the numerical procedure we determine the ground-state energy 
\begin{align}
E = \min_{C_{\underbar{n}}} \frac{\langle \Psi |H|\Psi\rangle}{\langle \Psi |\Psi\rangle} \equiv \langle H \rangle,
\end{align}
with a matrix-free approach: Using a variant of the power method \cite{Mises1929}, we iteratively rotate an initial state onto the ground state vector without having to construct the matrix explicitly. 
%We also separately compute the sum of kinetic and potential energy
%\begin{align}
%E_\mathrm{osc} = \langle H_{\mathrm{osc}}\rangle ,
%\end{align}
%and the interaction energy
%\begin{align}
%E_\mathrm{int} = E - (E_\mathrm{kin} + E_\mathrm{pot}) .
%\end{align}
Numerical computations are done using the NECI \cite{neci} software in deterministic mode. Even larger Hilbert spaces could be explored using stochastic algorithms for exact diagonalization such as 
Full Configuration Interaction Quantum Monte Carlo (FCIQMC) \cite{booth_fermion_2009}.

\section{%Resolution of the Gauss interaction potential 
Resolving the Gaussian pseudopotential\label{sec:resol}}

\subsection{Unscaled harmonic oscillator basis} \label{sec:unscaled}

\begin{figure*}
    \centering
    \includegraphics[scale=0.3]{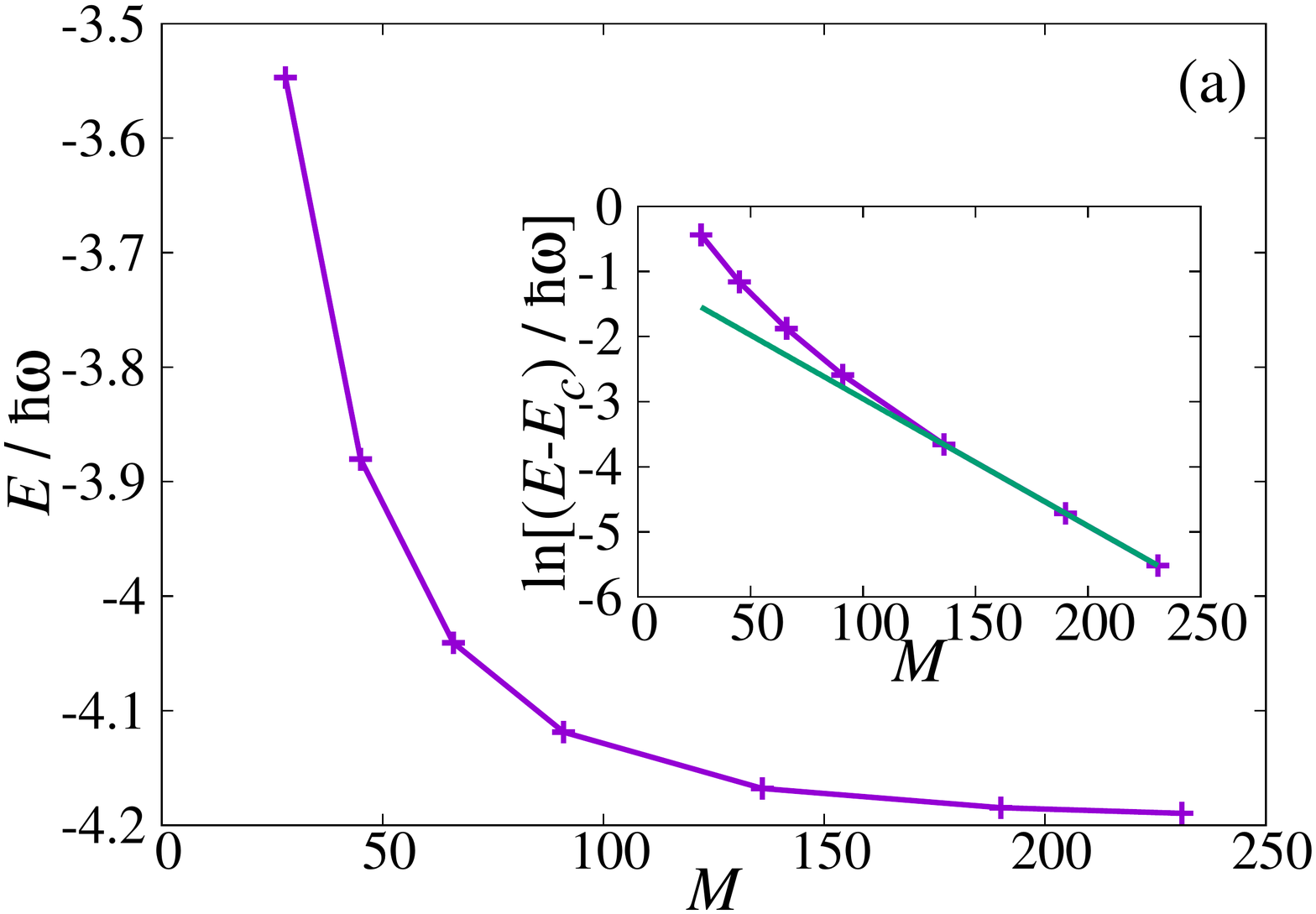}
    \includegraphics[scale=0.3]{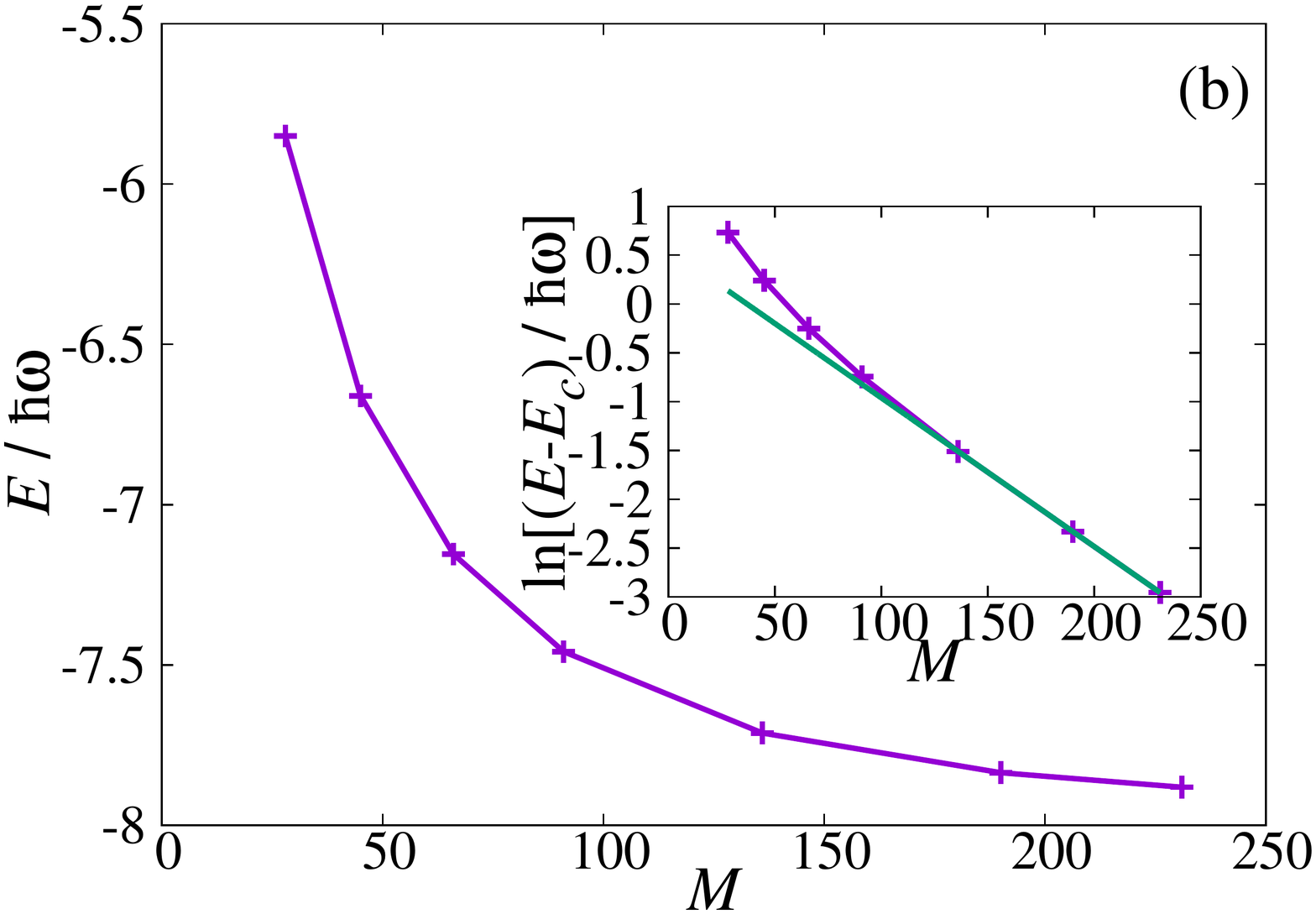}
    \includegraphics[scale=0.3]{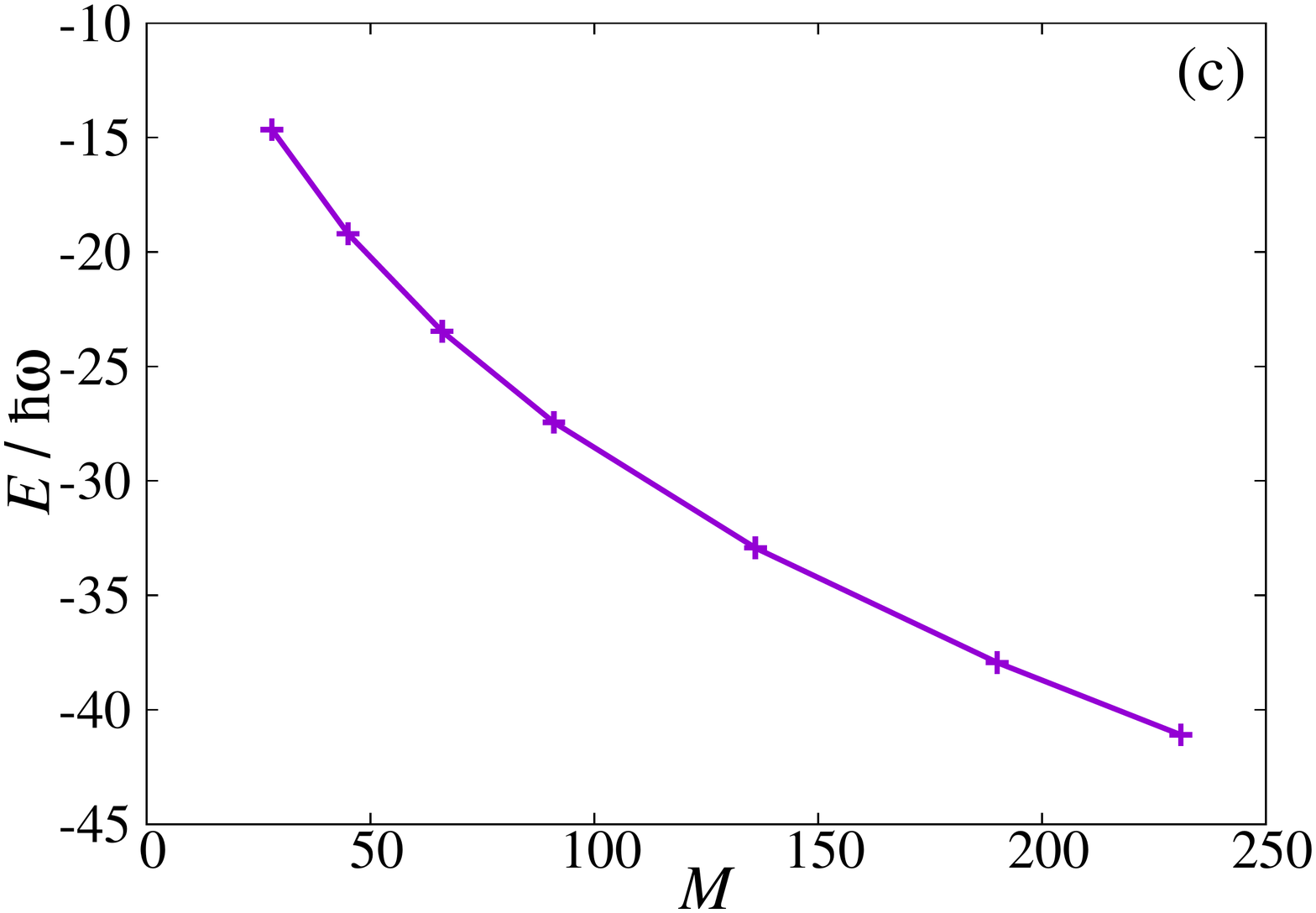}
    \includegraphics[scale=0.3]{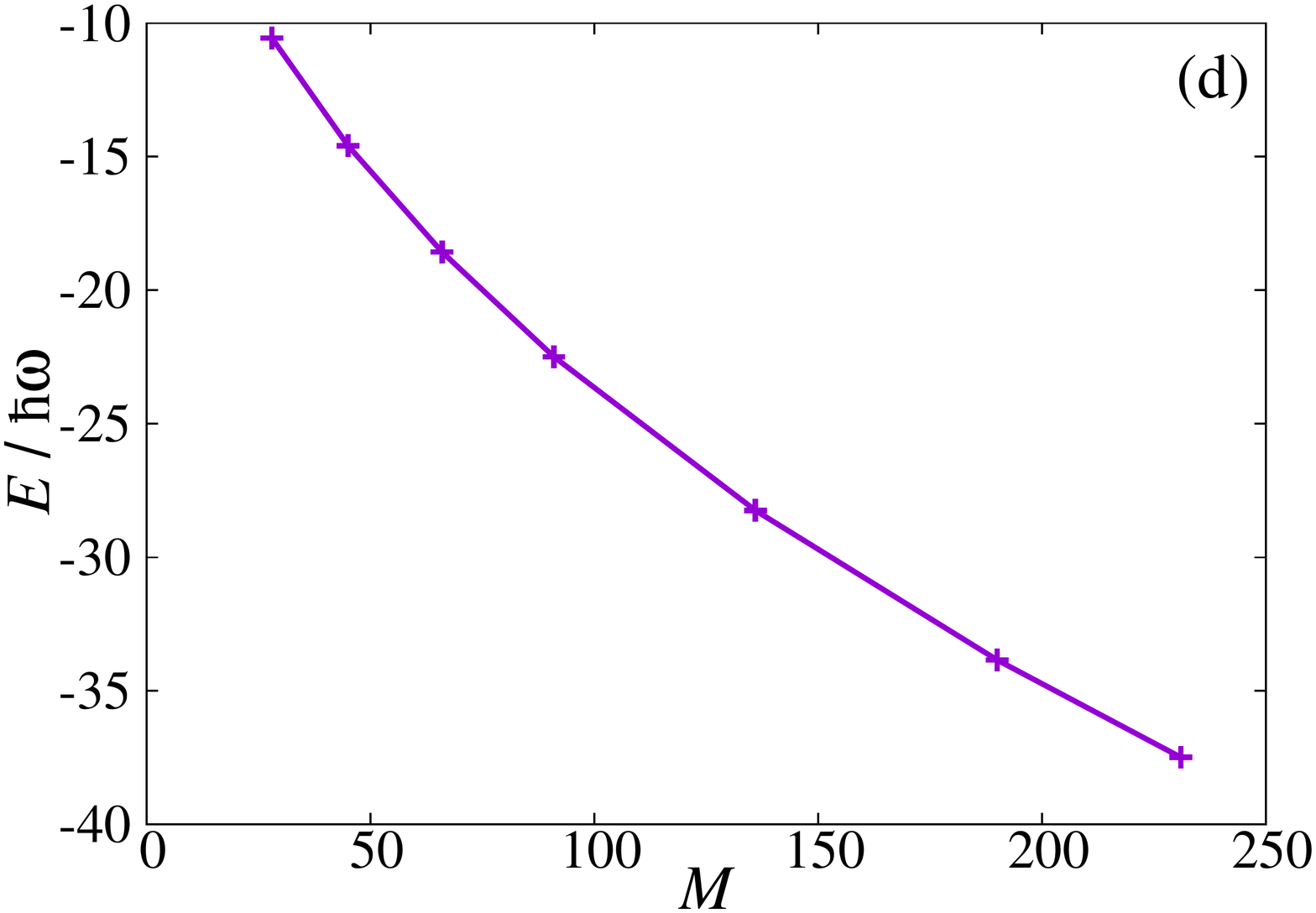}
 \caption{Convergence of the ground-state energy for three fermions with attractive Gaussian pseudopotential interactions in the unscaled harmonic oscillator basis. The energy from exact diagonalization in the finite multiparticle basis of Eq.~\eqref{eq:FockStates} is plotted vs  the size $M$ of the single-particle basis for Gaussian pseudopotentials of different widths $R$: (a) $R=l$,  (b) $R=0.8l$, (c) $R=0.3l$, and (d) $R=0.1l$, where $l=\sqrt{\hbar/m\omega}$ is the length scale of the harmonic trapping potential.  The insets in Figs. 1(a) and 1(b) show logarithmic plots of the same data as the main graph and demonstrate that a regime of exponential convergence is reached.
The extrapolated limiting values of the energy 
$E_c$ are obtained by nonlinear fitting of the exponential function $Ae^{-B M}+E_c$ to the last three data points: (a) $E_c=-4.19359\hbar \omega$ and (b) $E_c=-7.93323\hbar \omega$.
The interaction strength $\ln(l/a_s)=3.0$ is kept constant for all panels and the corresponding amplitude parameters 
$V_0$  are determined numerically following the procedure described in Ref.~\cite{jeszenszki_s_2018}: (a) $V_0=19.8237 \hbar \omega$, (b) $V_0=19.6329 \hbar \omega$, (c) $V_0=18.2369 \hbar \omega$, and (d) $V_0=13.024 \hbar \omega$.
%\todo[inline]{In the text we are claiming that the upper two panels reach the regime of exponential convergence. Can we demonstrate this explicitly with inset plots?\\
%Also: Please put label (a) to (d) on the panels.}
}
    \label{fig:convergencerate}
\end{figure*}

Figure \ref{fig:convergencerate} shows the ground state energy of three interacting fermions (two spin-up and one spin-down) according to the Hamiltonian \eqref{eq:Hfull} after full diagonalization with a finite Fock basis. We are using the Fock-Darwin form \eqref{eq:FDorbital} of the (unscaled) harmonic oscillator eigenfunctions of the single-particle Hamiltonian \eqref{Hoscdef} up to shell $n=20$, which yields up to $M=231$ single-particle basis functions. The maximal dimension of the computational Hilbert space 
%$\binom{M}{N_\uparrow}\binom{M}{N_\downarrow}$ 
for the three fermions with the zero total angular momentum slot  
is 
%thus $\approx 6 \times10^6$
$\approx 1.6 \times10^5$, which is 
already a significant size for  the deterministic diagonalization with available computational resources. 
%\todo[inline]{Please check numbers!}

It is clearly seen in Fig.~\ref{fig:convergencerate} that these basis set sizes are not sufficient to enter a regime of exponential convergence except for  Figs. 1(a) and 1(b), where this regime is reached for the last few data points as  seen from the insets. In these cases the width of the pseudopotential $R \approx l$ is close to the length scale of the trapping potential. Since the mean particle separation in the harmonic trap will also be of the same order $l$, or even smaller, this pseudopotential does not provide a useful approximation for the zero-range contact interactions that are relevant for modeling experiments with ultracold neutral atoms. As the lowest energy values reached for each pseudopotential width $R$ change significantly between the different panels of Fig.~\ref{fig:convergencerate}, it is also apparent that $R\ll l$ is a necessary condition for a useful, convergent approximation of the zero-range limit. Even without more sophisticated analysis, it is apparent from the results of Fig.~\ref{fig:convergencerate} that the necessary extrapolations the infinite basis set ($M\to\infty$) and zero-range ($R\to 0$) limit will be challenging to achieve.
%\begin{figure}
%  \centering
%  \input{as005L1.tex}
%  \caption{try}
%\end{figure} 

\subsection{Length scale resolution} \label{sec:length}

Since we are using a smooth Gaussian pseudopotential, we should expect that, for a sufficiently large basis set, the energy will converge exponentially to a limiting value with the size of the single-particle basis set. Indeed, it is well known that the Fourier transform of a Gaussian function yields again a Gaussian, which decays, in fact, faster than exponential in the tails. Sampling a Gaussian potential function in momentum space, should thus lead to at least exponential convergence, once the basis set is large enough to sample the tails of the Fourier-transformed Gaussian in momentum space. The necessary condition to reach this regime is that the basis set can resolve length scales that are smaller than the length scale $R$ of the Gaussian.

We now use this argument as a motivation to consider the length scale resolution of a two-dimensional harmonic oscillator basis. In order to keep the basis set independent from the Hamiltonian of Eq.~\eqref{Hoscdef}, we consider basis functions that are eigenfunctions of a harmonic oscillator with frequency $\tilde{\omega}$ and corresponding length scale $\tilde{l} = \sqrt{\hbar/m\tilde{\omega}}$. For a one-dimensional harmonic oscillator, the $p$th excited state has a spatial extent that can be estimated by the classical turning point $x_t$:
\begin{align}
 (p+\frac{1}{2})\hbar\tilde{\omega} = \frac{1}{2} m \tilde{\omega}^2 x_t^2,
\end{align}
or $x_t = \sqrt{2p+1}\, \tilde{l}$. The set of $M_{\mathrm{1D}} = p+1$ harmonic oscillator functions up to the $p$th excited states provides an approximately homogeneous sampling of the interval $[-x_t, x_t]$ at a length scale 
\begin{align}
l_\mathrm{res}=\frac{2x_t}{M_{\mathrm{1D}}} =  \frac{2\sqrt{2p+1}}{p+1}\tilde{l}. \label{resolfrist}
\end{align}
In order to connect this result to the number $M$ of  two-dimensional harmonic oscillator basis functions, we construct the latter as a product basis (of Hermite functions) with an energy cutoff. This yields
\begin{align}
M&=\sum_{i,j=0}^{i+j \le p} 1 = \frac{(p+1)(p+2)}{2} \ . \label{Mandp} 
\end{align}
We can thus relate the resolution length scale $l_\mathrm{res}$ to the size of the basis and obtain
\begin{align}
l_{\mathrm{res}} &= \frac{4 \sqrt{\sqrt{8M+1}-2}}{\sqrt{8M+1}-1} \tilde{l} , \label{landM}
\end{align}
which can be solved for $M$ to yield
\begin{align}
M \approx 32  \left( \frac{\tilde{l}}{l_\mathrm{res}} \right)^4  , \label{Mapproxeq}
\end{align}
where lower order terms were neglected assuming $\tilde{l} \gg l_\mathrm{res}$. 

Equation~\eqref{Mapproxeq} provides an estimate for the size of the single-particle basis needed to resolve a length scale $l_\mathrm{res}$. For the situation of Sec.~\ref{sec:unscaled} where $\tilde{l}=l$ we can estimate the minimum size of the basis set to be able to resolve the pseudopotential length scale $R$ as
\begin{align} \label{eq:Mmin}
M_\mathrm{min} \approx  32 \left( \frac{l}{R} \right)^4 ,
\end{align}
i.e.~the required basis set size increases rapidly when the length scale $R$ (and thus the range) of the pseudopotential is decreased. 

Specifically, it means that reaching an exponentially convergent regime should be quite achievable when the pseudopotential width is of the same order of magnitude as the oscillator length $l$. For $R = l$ and $R=0.8 l$ we would require a minimum of 32 and 78 single-particle basis functions, respectively, which means about 16,000 and 200,000 Fock states for three fermions.
This is consistent with the numerical results of Fig.~\ref{fig:convergencerate} obtained with up to $M=230$ single-particle basis functions.

In order to explore the physics of short-range interactions, however, we may need to use narrower pseudopotentials. With modest choices of $R=0.3 l$ and $R=0.1l$ the number of required single-particle basis functions already increases to about  4,000 and 300,000, respectively, corresponding to about $3 \times 10^{10}$ and $1 \times 10^{16}$ multiparticle basis functions. In this case, even the storage of the wave function is very expensive as it would require 0.2 and 100,000 terabytes of memory, respectively. Although exploiting symmetries, using approximation schemes, or stochastic sampling techniques can reduce these requirements \cite{booth_fermion_2009,christensson_effective-interaction_2009}, the quartic scaling in Eq.~\eqref{Mapproxeq} does not seem pleasant.

\subsection{Scaled harmonic oscillator basis}\label{sec:scaled}

%Setting the parameter $l_u$ to be equal to the unit length of the harmonic oscillator $l$, the single particle basis becomes the solution of the noninteracting system. 
%It has an advantage as it already includes the significant effects of the external trapping potential. Though, the eigenfunctions with higher single-particle energy extend to further region from the origin, where the actual density of the physical system is negligible. Moreover, it also has a difficulty to describe the short-range correlations, which have a more pronounced effects in the strongly interactive regime. Hence In order to give an optimal choice of parameter $l_u$ let us 

The required size of the single-particle basis can be decreased by introducing an appropriate scaling of the basis functions. The main idea is the following: Increasing the number of basis functions not only improves the resolution length scale $l_\mathrm{res}$ (by making it smaller) but also increases the largest length scale that can be described by the basis, which is given by $2 x_t = 2\sqrt{2p+1}\;\tilde{l}$, i.e., through the classical turning point. 
%Since the largest length scale that needs to be resolved by the single-particle basis is independent of the basis s
This means that the basis functions can be scaled according to the basis set size while still resolving the largest length scale of the problem (e.g., twice the Thomas-Fermi radius of a cold atomic cloud). Let us denote this largest relevant length scale as $\gamma l$, where $\gamma$ is the dimensionless form of this length scale in units of the length scale $l$ of the trapping potential.  Note that $\gamma$ is determined by the physical properties of the system (i.e., particle number, interaction strength, etc.) and is thus independent of the basis set size. For the required basis set length scale we then obtain
\begin{align}
\tilde{l}=\frac{\gamma}{2\sqrt{2p+1}} l \ . \label{luandl}
\end{align}
With Eq.~\eqref{resolfrist} this yields $l_\mathrm{res}= \gamma l/(p+1)$ and using Eq.~\eqref{Mandp} to solve for $M$ we obtain
\begin{eqnarray}
M = \frac{\gamma^2}{2} \left( \frac{l}{l_\mathrm{res}} \right)^2 + \frac{\gamma}{2}  \frac{l}{l_\mathrm{res}} \ .
\end{eqnarray}

Applying this result to the problem of resolving a Gaussian pseudopotential with length scale $l_\mathrm{res} = R$ and taking the leading term for $l\gg R$, we obtain the revised relation for the minimum size of the single-particle basis,
\begin{align} \label{eq:MminScaled}
M_\mathrm{min} \approx  \frac{\gamma^2}{2} \left( \frac{l}{R} \right)^2 ,
\end{align}
when the length scale of the single-particle basis is optimally scaled with the size of the basis set. Compared to the result \eqref{eq:Mmin} for the unscaled basis, the power-law scaling of the required basis set size with the pseudopotential length scale $R$ in Eq.~\eqref{eq:MminScaled} is improved by two orders.

\subsection{Estimating the interaction-dependent prefactor}\label{sec:prefactor}

As $\gamma l$ represents the largest length scale that has to be resolved by the single-particle basis, i.e., the maximal spatial extent of the system, the factor $\gamma$ depends on the details of the Hamiltonian, which, for our example, are the particle number content and the strength of the contact interaction.
Although the exact value of $\gamma$ is difficult to calculate we can estimate its value and present upper and lower bounds from limiting cases that are simple to analyze.

Specifically for fermions with attractive short-range interactions as per Eq.~\eqref{eq:Hfull}, the size of the trapped noninteracting gas cloud will provide an upper bound, while a lower bound can be obtained from the strongly interacting limit where fermion pairs form deeply bound composite bosons while excess fermions remain with little residual interactions.

Starting with the upper bound, we consider a noninteracting Fermi gas with a possibly unequal population of spin-up and spin-down fermions. The largest length scale is determined by the Fermi pressure of the majority component with the quantum number $p^{\mathrm{majority}}$ along a single spatial dimension
% directions on the Fermi surface,
%with particle number  $N_\mathrm{majority}$ and `effective' one-dimensional quantum number $p^{\mathrm{majority}}$: 
\begin{eqnarray}
\gamma_\mathrm{upper} &=& 2\sqrt{2p^{\mathrm{majority}}+1} \  . \label{gammaest} %\\
%p^\mathrm{majority} &=& p_x^\mathrm{majority} +p_y^\mathrm{majority} \ ,%2n_\mathrm{Fermi} + \ell_\mathrm{Fermi} = \frac{2\epsilon_\mathrm{Fermi}}{\hbar \omega}-1 \ , \nonumber
\end{eqnarray}
%
%Although we do not know the exact value of $\gamma$, we can give an upper and lower bound for the attractive interaction  
%%and the repulsive interactions cases 
%by examining the non-interacting and strongly interacting limits.  In the non-interacting ground state 
%the spin-up and spin-down particles independently fills up the single-particle orbitals with the lowest single particle energy. The largest length scale is determined by the most delocalized single-particle orbital, which also has the largest single-particle energy. As the attractive interaction decreases the average particle-particle distance, this length scale also gives an upper bound for the largest length scale of the system. Transforming the occupied single-particle orbitals from the majority component to the localized ones, we can express a relation between the largest length scale, $\gamma_\mathrm{upper}$ 
%and the effective one-dimensional quantum number, $p^{\mathrm{majority}}$,
%\begin{eqnarray}
%\gamma_\mathrm{upper} &=& \sqrt{2p^{\mathrm{majority}}+1} \  , \label{gammaest} \\
%p^\mathrm{majority} &=& p_x^\mathrm{majority} +p_y^\mathrm{majority} \ .%2n_\mathrm{Fermi} + \ell_\mathrm{Fermi} = \frac{2\epsilon_\mathrm{Fermi}}{\hbar \omega}-1 \ , \nonumber
%\end{eqnarray}
%where 
%$\epsilon_\mathrm{Fermi}$ is the highest single-particle energy (Fermi energy),
%$p_x^\mathrm{majority}$, $p_y^\mathrm{majority}$ are the quantum numbers for each spatial directions. 
%\todo[inline]{Why do we sum the two components? Shouldn't the ``one-dimensional quantum number'' just be the larger of the two?}
The parameter $p^\mathrm{majority}$ can be expressed in terms of particle number $N_\mathrm{majority}$ by considering \rref{Mandp} and using the information that only one particle can occupy one spatial orbital from the same spin component to yield
\begin{eqnarray}
\gamma_\mathrm{upper} &=& 2\sqrt{2\ceil*{\frac{\sqrt{8N_\mathrm{majority}+1}-3}{2}}+1} \ , \label{gammaupper}
\end{eqnarray}
where $\ceil*{x}$ is the ceiling function \cite{eric_w._weisstein_ceiling_2018}.

For strong short-range attractive interactions, pairs of spin-up and spin-down particles form  point like effective bosons \cite{madison_strongly_2015,petrov_superfluid_2003}. The interactions between the effective bosons are repulsive but vanish in the limit of strong attraction between fermions.
In this limit we thus obtain a noninteracting system where all of the bosons 
occupy the lowest single-particle orbital. In the spin-balanced system, the largest length 
scale will thus be given by the length scale of the harmonic oscillator trapping potential. This length scale provides a lower 
bound of the largest length scale of the system as the effective repulsive interactions can only 
increase the average particle-particle distance. We thus obtain a lower bound of
\begin{eqnarray}
\gamma^{\mathrm{sb}}_\mathrm{lower} &=& 2 \ , 
\label{gammalowersb}
\end{eqnarray}
where the index `$\mathrm{sb}$' stands for the spin balanced case.

In the spin-imbalanced case, excess fermions from the majority component that are not locked-up in effective bosons who will maintain Fermi pressure. Indeed, the excess fermions have weak repulsive interactions with the effective bosons that also  vanish in
the strongly interacting limit (regarding the original interactions between fermions) \cite{madison_strongly_2015,petrov_superfluid_2003}. Hence, the lower bound for the largest length scale 
can be tightened by considering a noninteracting Fermi gas of the excess fermions following the same argument as above. We obtain
\begin{eqnarray}
\gamma^{\mathrm{si}}_\mathrm{lower} = 
2\sqrt{2\ceil*{\frac{\sqrt{8 \left|N_\uparrow - N_\downarrow \right|+1}-3}{2}}+1} \ , \label{gammalowersi} 
\end{eqnarray}
where the index "$\mathrm{si}$" stands for the spin-imbalanced case.

We can see from  \rrefsa{gammaupper} and \rrefsb{gammalowersi} that the largest relevant length scale $\gamma l$ increases with particle number. To leading order the bounds become 
\begin{eqnarray}
\gamma_\mathrm{upper} &\approx& 2\sqrt[4]{8N_\mathrm{majority}} 
\ , \label{approxgammanoint} \\
 \gamma_\mathrm{lower}^{\mathrm{si}} &\approx& 2\sqrt[4]{8 \left|N_\uparrow - N_\downarrow \right|} \ .
 \label{approxgammalower}
\end{eqnarray}
Comparing with expression \eqref{eq:MminScaled} for the minimum size of the single-particle basis, we see that $M_\mathrm{min}$ increases with the square root of the particle number (that is  $N_\mathrm{majority}$ and 
$|N_\uparrow-N_\downarrow|$, respectively), i.e., requiring a larger single-particle basis for a larger particle number. We also see that the largest interaction dependence of  $M_\mathrm{min}$ can be expected in the spin-balanced case, whereas  $M_\mathrm{min}$ will approximately remain independent of interactions for large spin polarization (e.g., polaron physics), where  $N_\mathrm{majority} \approx |N_\uparrow-N_\downarrow|$.

%As it can be seen form \rrefsa{gammaupper} and \rrefsb{gammalowersi} the particle number increases the largest length scale of the system, which make harder the accurate description.
%Let us simplify the relation between the bounds of $\gamma$ and number of the particles 
%by considering the limit of large particle numbers and the large differences of number of particles between the different spin components in \rrefsa{gammaupper} and \rrefsb{gammalowersb},  
%\begin{eqnarray}
%\gamma_\mathrm{upper} &\approx& \sqrt[4]{8N_\mathrm{majority}} 
%\ , \label{approxgammanoint} \\
% \gamma_\mathrm{lower} &\approx& \sqrt[4]{8 \left|N_\uparrow - N_\downarrow \right|} \ .
% \label{approxgammalower}
%\end{eqnarray}
%Comparing the approximate equations \rrefsb{approxgammanoint} and \rrefsb{approxgammalower} with the quadratic dependence of $M$ on $\gamma$ in \rref{Mapproxgamma}, we find that $M$ increases with the square root of $N_\mathrm{majority}$ and 
%$|N_\uparrow-N_\downarrow|$ depending on the interaction regime. It also shows we can expect the largest changes of the radius at the spin-balanced case, while at large spin polarization the radius more or less constant as $N_\mathrm{majority} \approx |N_\uparrow-N_\downarrow|$.  

For large particle numbers the large length scale $\gamma l$ is well approximated by Thomas-Fermi theory as  $\gamma l= 2 R_\mathrm{TF}$, where $R_\mathrm{TF}$ is the Thomas-Fermi radius.
In Thomas-Fermi theory, the single-particle density $n(r)$ is found from solving
\begin{align} \label{TFapprox}
\mu_0 = V_\mathrm{ext}(r) + \mu[n(r)] ,
\end{align}
where  $V_\mathrm{ext}(r) =  \frac{1}{2}m \omega^2 r^2$ is the external potential, $\mu[n(r)]$ represents the chemical potential at local equilibrium [from the equation of state $\mu(n)$ of the homogeneous gas], and the constant $\mu_0$ is the chemical potential of the finite system \cite{pethick_bose-einstein_2002,giorgini_theory_2008,ketterle_making_2008}. The Thomas-Fermi radius $R_\mathrm{TF}$ is then the value of $r$ where $n(r)$ drops to zero.

For our case of a two-dimensional BCS mean-field theory yields a Thomas-Fermi radius $R_\mathrm{TF}$ that is independent of particle interactions \cite{Randeria1989}, i.e., the result \eqref{approxgammanoint}. In the strongly interacting limit of the balanced Fermi gas, we may use the known asymptotic form of the equation of state for a two-dimensional gas of bosonic dimers \cite{Schick1971,cherny_self-consistent_2004}
\begin{align}
\mu(n) = -\frac{2 \pi \hbar^2}{m} \frac{n}{\ln \left( a_{dd}^2 n \right)} ,
\end{align}
where the scattering length of bosonic dimers $a_{dd}$ is approximately $a_{dd} \approx 0.55(4) a_s$ \cite{bertaina_bcs-bec_2011,petrov_superfluid_2003}.
%
%In the strongly attractive limit the 
%interaction dependence of the largest length
%scale can be estimated through the 
%Thomas-Fermi radius for the spin-balanced case. 
%In the limit, where the particle-particle interaction dominates over the kinetic energy of the condensate, the well-known Thomas-Fermi equation \cite{giorgini_theory_2008,ketterle_making_2008} can
%be applied,
%\begin{eqnarray}
%\mu_0(r) &=& \frac{m \omega^2 r^2}{2}-\frac{2 \pi \hbar^2}{m} \frac{n(r)}{\ln \left[ a_{dd}^2 n(r) \right]} \ , \label{TFapprox}
%\end{eqnarray}
%where $\mu_0$ is the chemical potential, $n(r)$ is 
%the density of the bosonic dimers and 
%$a_{dd}$ is the bosonic dimer-dimer $s$-wave scattering length. The value of $a_{dd}$ can be well approximated with the 
%$s$-wave scattering length of the attractive fermions, $a_s$ \cite{bertaina_bcs-bec_2011}, 
%\begin{eqnarray*}
%a_{dd} & \approx & 0.55(4) a_s \ . 
%\end{eqnarray*}
The Thomas-Fermi radius
%, $R_\mathrm{TF}$, 
%is defined by the zero value of the density. Its approximate expression in the large particle number limit 
can be determined from \rref{TFapprox}. After some algebraic manipulations we obtain
\begin{eqnarray*}
R_\mathrm{TF}
&\approx& \sqrt{a_{dd}^2 N_d + \frac{4 \pi l^4 }{a_{dd}^2} 
\ln\left( \frac{8 \pi l^4e^{\gamma_\mathrm{E}+1}}{N_d a_{dd}^4} \right) } \ ,
\end{eqnarray*}
where $N_d$ is the number of the bosonic dimers and 
$\gamma_\mathrm{E}$ is the Euler-Mascheroni constant. The resulting approximation for the scaling factor $\gamma$ is $\gamma = 2 R_\mathrm{TF}/l$.

\begin{figure}
    \centering
    \includegraphics[scale=0.3]{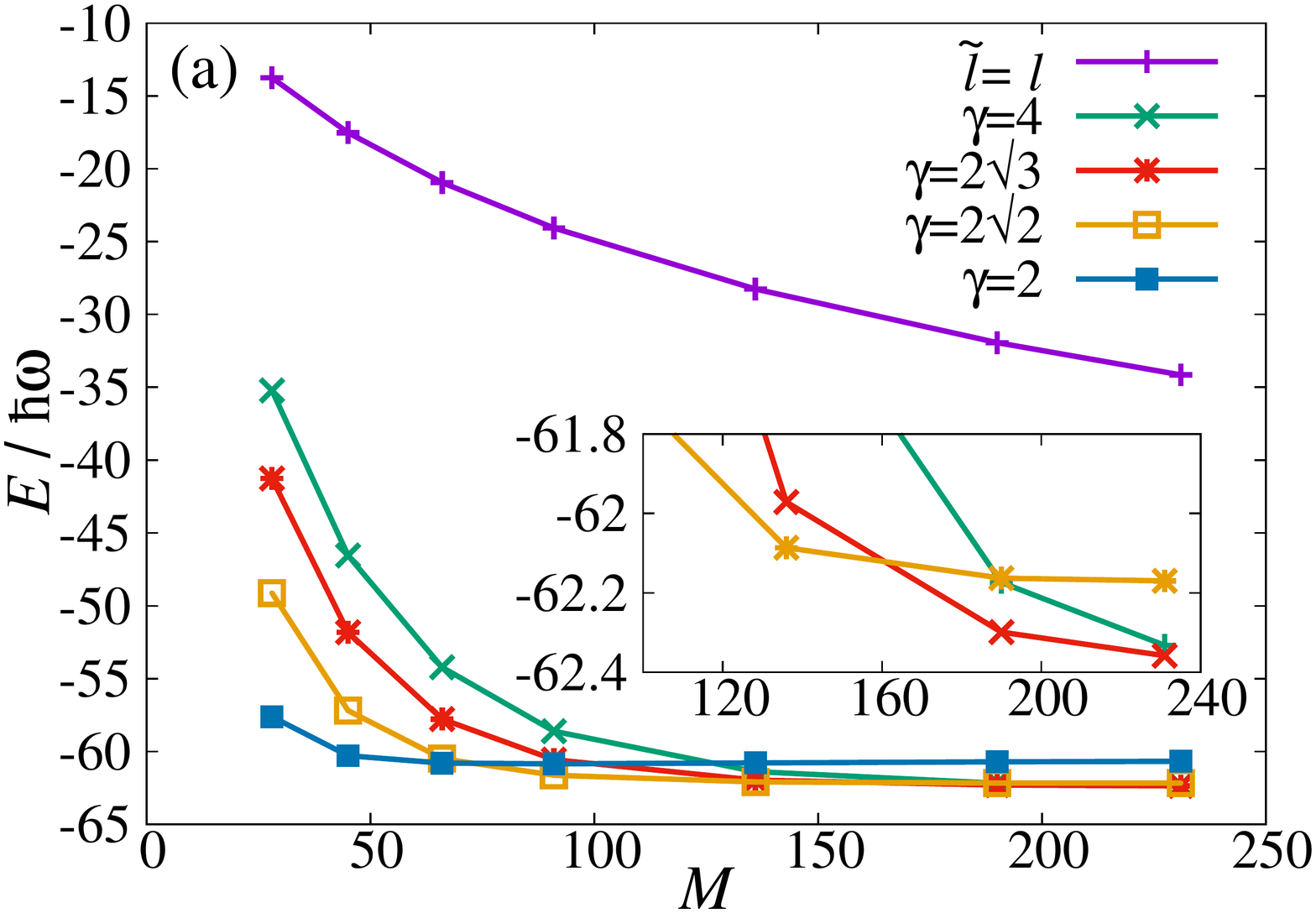}
    \includegraphics[scale=0.3]{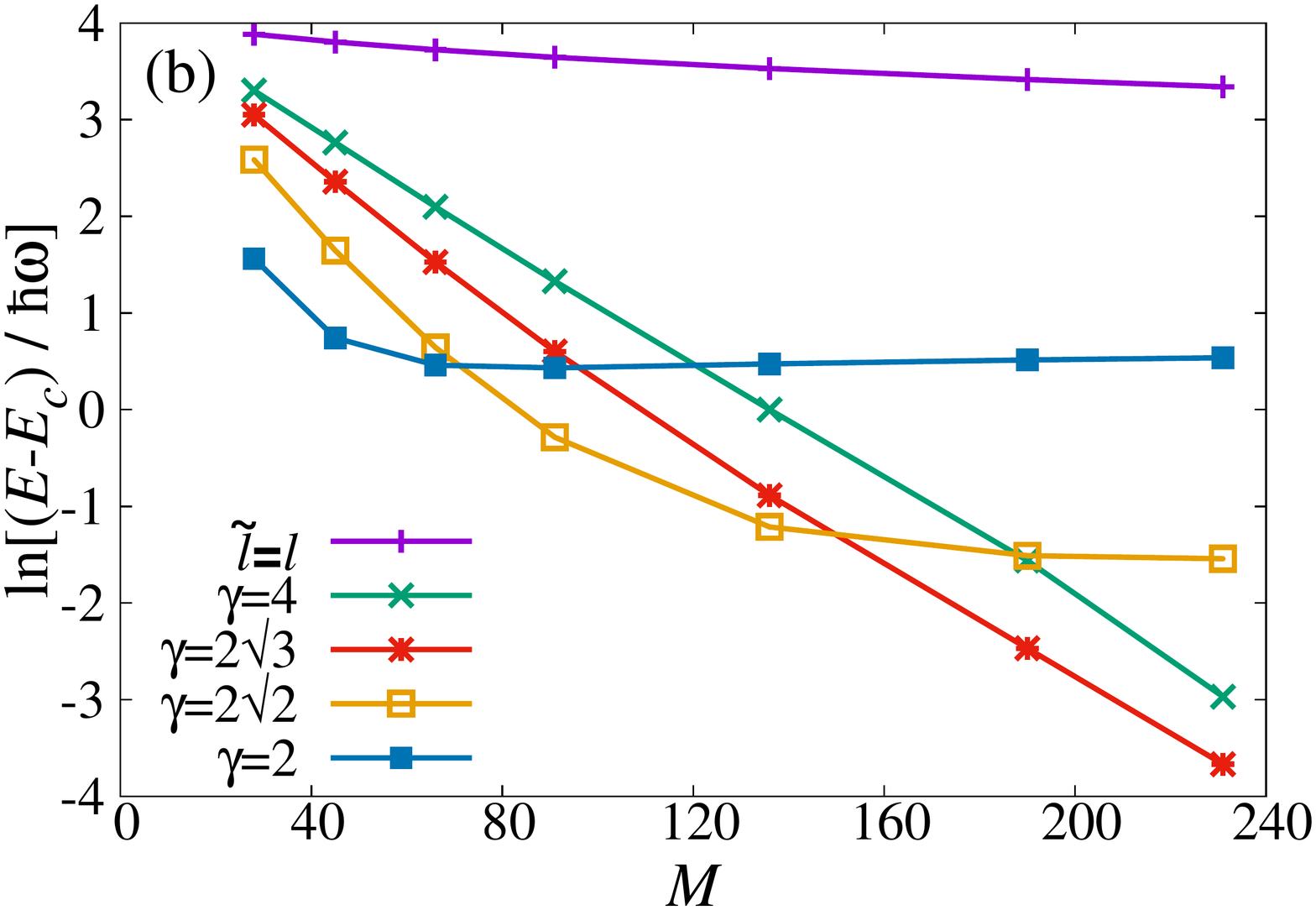}
\caption{Energy convergence in a scaled harmonic oscillator basis at $R=0.3l$  with different values of the factor $\gamma$. 
For reference, we also show results for the unscaled basis (green crosses) identical to Fig.~\ref{fig:convergencerate}(c).
(b) A logarithmic representation of the same data as in (a).  
%The curve, where the $\tilde{l}$ parameter is set to $l$, is also plotted for reference.
The extrapolated limiting value of the energy $E_c$ is obtained by nonlinear fitting of the exponential function $Ae^{-B M}+E_c$ to the last three data points of the $\gamma=2\sqrt{3}$ data with $E_c=-62.38392715\hbar \omega$.
 $V_0=18.2369 \hbar \omega l^2$ corresponding to  $\ln(l/a_s)=3.0$.
%The potential strength $V_0$  is determined with a numerical procedure described in Ref. \cite{jeszenszki_s_2018}, in order to obtain the $\ln(l/a_s)=3.0$ interaction strength. 
%    \todo[inline]{Captions will need to be updated.}
    }
    \label{fig:scaling}
\end{figure}

In order to test the prediction of Eq.~\eqref{eq:MminScaled} for reduced requirements for the size of a scaled single-particle basis we consider again the example of two spin-up particles and one spin-down particle with attractive interactions. The upper and lower bounds for $\gamma$ can be easily calculated from \rrefsa{gammaupper} and \rrefsb{gammalowersi} to give $\gamma_\mathrm{upper} = 2\sqrt{3}$ and $\gamma_\mathrm{lower} = 2$. The required minimum number of basis functions $M_\mathrm{min}$ from Eq.~\eqref{eq:MminScaled} then yields 20 and 60, respectively, for $R=0.3l$ [reduced from $M_\mathrm{min} \approx$ 4,000 for the unscaled basis according to Eq.~\eqref{eq:Mmin}].
%
%In order to test the numerical improvement of the scaling factor, let us examine the case of two spin-up particles and one spin-down particle with exact diagonalization. 
%The bounds of $\gamma$ with 
%$\gamma_\mathrm{upper} = \sqrt{3}$ and $\gamma^\mathrm{att}_\mathrm{lower} = 1 $
%can be easily calculated from \rrefsa{gammaupper} and \rrefsb{gammalowersb}.
%%in the non-interacting limit, one of the spin-up particles occupies the first single-particle excited states. In that case $p_\mathrm{Fermi}=1$, which means an upper bound $\gamma_{\mathrm{upper}}=\sqrt{3}$. In a strongly attractive limit, one of the spin-up and the spin-down fermion form a pair and occupies the lowest single-particle state with the remaining spin-up fermion. It means that the lower bound $\gamma_{\mathrm{lower}}=1$. 
%The improvement is significant for the narrow Gaussian widths. The required number of basis functions decreases between 20 and 60 from 4000 for $R=0.3l$.  
Similarly for $R=0.1l$ the number decreases from 300,000  to 200--600 basis functions. 
In Fig.~\ref{fig:scaling} we show the ground-state energy  $R=0.3l$ with different scaling factors $\gamma$. The results demonstrate that the regime of exponential convergence can be reached  with the scaled basis, even though it was unattainable with the unscaled basis with the available computational resources. 
The smallest scaling factors of $\gamma=2$ and $\gamma=2\sqrt{2}$ are seen to result in an (unphysical) energy bias. This can be understood by the fact that the lower bound $\gamma=2$ and $\gamma=2\sqrt{2}$ underestimate the system size and thus the scaled basis set does not cover the whole area occupied by finite particle density. Increasing the value of $\gamma$ from the lower bound eliminates the bias but also eventually leads to a reduced convergence rate. At the upper bound  ($\gamma=2\sqrt{3}$) the computation is seemingly free of bias but convergence of the energy is greatly improved compared to the unscaled basis set. Hence, we find that the upper bound can safely be used for the accurate determination of the ground-state energy.

Extrapolating our results to larger particle numbers, we need to consider the following issues: First, adding more particles at constant $M$ increases the size of Hilbert space (and thus computational complexity) due to the binomial scaling with $N$ and $M$. Whether the size $M$ of the single-particle basis has to be changed depends on how the relevant length scales change. The smallest length scale only depends on the choice of the pseudopotential and is invariant with particle number. Thus the required $M$ for the unscaled harmonic oscillator basis is independent of particle number (as long as $M$ is larger than the majority-component particle number). For the scaled-basis approach we have to consider that the relevant largest length scale may change. For example, adding another particle in the minority component (say going from three to four fermions where two are spin-up and two are spin-down) will compress the wave function towards the center of the trap for attractive interactions and thus reduce the size of the relative wave function. Also the center-of-mass wave function will shrink due to the larger total mass.  
%with one spin-up and/or two spin-down particles, the attractive interaction further compresses the particles to the center of the trap. 
Therefore, an even smaller value of $\gamma$ can be applied leading to smaller $M_\mathrm{min}$. However, if the number of the particles in the majority component exceeds three particles, one of the particles in the noninteracting case occupies the next shell, which increases the size of the system and  the upper bound for $\gamma$ increases to $\gamma_\mathrm{upper} = 2\sqrt{5}$.
In this case, the number of  single-particle basis functions increases to $M_\mathrm{min} \approx 120$, which is still much less than the required number of the unscaled basis functions, 
$M_\mathrm{min} \approx 4000$. The increment of $M_{\mathrm{min}}$ is less severe for larger particle numbers, again due to the effect of attractive interactions. %The $\gamma_\mathrm{upper}$ increases to  $2\sqrt{7}$ at $N_\mathrm{majorty}=7$ leading to $M_\mathrm{min} \approx 160$. 
Using $M_\mathrm{min} \approx 250$ scaled single-particle basis functions, we can describe about 15 particles in the majority component, which can be 30 particles in the spin-balanced case. For repulsive interaction, the relevant largest length scale of the system, the Thomas-Fermi radius, will generally increase with interaction strength leading to a larger scaling factor $\gamma$. 

\begin{figure}
    \centering
    \includegraphics[scale=0.3]{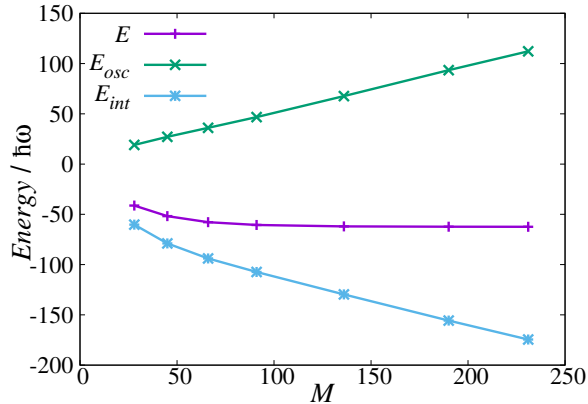}
\caption{Convergence of energy components with respect to increasing size of the single-particle basis set $M$.
While the total energy $E = E_\mathrm{osc} + E_\mathrm{int}$ converges nicely with increasing  $M$ at  $R=0.3l$  and  $\gamma=2\sqrt{3}$ (data from Fig.\ \ref{fig:scaling}), the single-particle part of the energy $E_\mathrm{osc}$ and the interaction energy $E_\mathrm{int}$, which more sensitively probe short-range correlations in the wave function, have not yet separately converged within the accessible range of $M$ values.
}
    \label{fig:intenerg}
\end{figure}

Finally, we would like to illustrate that accessing the regime of
exponential convergence is significantly easier
for the total energy than for other physical quantities that more sensitively probe the short-range correlations in the wave function.
Let us examine the energy partition $E = E_\mathrm{osc} + E_\mathrm{int}$, where $E_\mathrm{osc} = \langle H_{\mathrm{osc}}\rangle$ collects the single-particle contributions of kinetic and potential energy [see Eq.\ \eqref{eq:Hfull}] ,
%and thereby probes the single-particle density matrix 
and the interaction energy $E_\mathrm{int}$ probes the two-particle density matrix. While both quantities reach exponential convergence for $R=l$ and $R=0.8l$ in a similar way as the total energy (data not shown), they do not converge for the narrower Gaussians within the available range of $M$. Figure \ref{fig:intenerg} demonstrates this behavior for $R=0.3l$ where the total energy seemingly converges and does not significantly change for $M>100$ on the scale of the plot,
% (same data as in Fig.~\ref{fig:scaling}), 
but the single-particle and interaction energy parts still show a strong $M$-dependence within the accessible range of basis set sizes. 
While $E_\mathrm{osc}$ and $E_\mathrm{int}$ serve as useful and sensitive measures of low-order correlations for finite-size Gaussian pseudopotentials, please note that they lose meaning  in the zero range limit $R\to 0$ where they separately diverge. Only the sum (i.e., the total energy) remains meaningful, which is a known feature of zero-range pseudopotentials in two and three dimensions \cite{Busch1998}. 

%
% two energy partitions, the single-particle contributions, which equals to the sum of the kinetic energy and the external trapping potential energy, 
%and two-particle contributions, which gives the particle-particle interaction energy. 
%In Fig.~\ref{fig:intenerg}, the ground-state energy is plotted together with the energy partitions at different number of single-particle 
%basis functions.  
%For the energy partitions the convergence cannot be achieved even at the largest accessible single-particle basis set, whereas the ground-state energy does not significantly change after $M=100$. 
%
%We also expect similar slower convergence behavior for the correlation functions as well. In the two-particle
%correlation function near to the particle-particle coalescence the Gauss pseudopotential indicates a change,
%which is related to $R$. This error is about the same order as the error in the wave function, which also
%propagates to the energy partitions after integrating out to the correlation functions.
%Calculating the total energy, the first order error from the kinetic energy and the particle-particle energy eliminates each other providing errors only in the second order \cite{helgaker_molecular_2008}.
%These small demonstrative examples further encourage us to develop an efficient numerical algorithm, with 
%which the convergence behavior can be accelerated. 

%The convergence of the energy has been plotted for $R=0.3l$ in Fig. \ref{fig:scaling} for different values of $\gamma$. As it can be seen all the values between the upper and 
%lower bound significantly improve the convergence compare to the unscaled basis. Choosing the lower bound of $\gamma$ with $\gamma=1$ the bias is present due to the poor description of the region, which further from the origin. Applying larger $\gamma$ the speed of the convergence decreases at a smaller basis, but it eliminates the bias originated from the boundaries of the system. The upper bound ($\gamma=\sqrt{3}$) seemingly free from any bias, but still significantly improve the convergence rate. Hence, at least in this example the upper bound safely can be used for the determination of the converged energy.
%

\section{Conclusion and outlook}\label{sec:concl}

In this work we considered the question whether smooth pseudopotentials are a good choice for representing short-range interactions in numerical approaches that rely on a Fock-state basis constructed from a finite set of single-particle functions. The combination of smooth pseudopotentials and an asymptotically complete set of (smooth) single-particle basis functions promises exponential convergence. This regime of exponential convergence can only be reached, however, if the basis set is large enough to resolve the relevant the length scales of the problem.

In order to isolate the effects of the single-particle basis we have used an exact diagonalization procedure to capture the many-particle quantum physics.
For an example system of experimental interest (a two-dimensional harmonic trapping potential with attractively-interacting fermions) we have derived simple expressions for the required minimum size of the single-particle basis in order to resolve a given pseudopotential length scale. The algebraic scaling of $l_\mathrm{res}^{1/4}$ can be improved to $l_\mathrm{res}^{1/2}$ by scaling the basis set but remains algebraic with the required resolution length scale. An additional algebraic scaling of the required basis set size is found to apply to the particle number. With numerical example calculations for three particles we could demonstrate that the exponential convergence regime could indeed be reached, albeit in a regime where the pseudopotential length scale is not much smaller than the particle separation of length scale of the trapping potential.

In order to faithfully represent the physics of short-range interacting particles, as is relevant for ultracold quantum gases of neutral atoms, it would be necessary to reduce the length scale of the pseudopotential much further and extrapolate to the zero range limit. Although such extrapolation has been demonstrated in few-body systems using different computational approaches \cite{blume_few-body_2012}, it does not seem feasible for the current approach. Looking towards the treatment of larger particle numbers and simultaneous extrapolation to the zero-range limit, one has to consider that the benefits of asymptotically exponential convergence of the single-particle basis are counteracted by the algebraically scaling requirements for the minimum size of the single-particle basis, both with particle number and with the ratio of extremal length scales that have to be resolved. 

It would be interesting to examine the basis set convergence in Jacobi coordinates, where one, typically large, length scale can be removed by separating out the center-of-mass motion.
The center-of-mass length scale can become the largest length scale in the system at 
strong attractive interaction  and weak trapping (e.g., for low-mass bright solitons or droplets),
and hence its elimination from the numerical calculations should accelerate the convergence properties according to the arguments discussed in Sec.\ \ref{sec:resol}. 
In addition, the overall computational complexity is reduced by eliminating the center-of-mass degrees of freedom.
Calculations in Jacobi coordinates have been  performed, e.g., for three particles in one dimension \cite{harshman_symmetries_2012,garcia-march_distinguishability_2014}. Extensions to higher dimensions and larger particle numbers are complicated by the fact that fermionic or bosonic permutation symmetries manifest themselves by more complex symmetries in Jacobi coordinates that are more difficult to treat \cite{harshman_symmetries_2012}.

%By extending this method to two dimensions we can examine the convergence properties of the non-trivial relative motion part of the wave function. 

A remaining alternative approach is to replace the smooth pseudopotential by a renormalized contact interaction. This has the advantage of removing the artificial length scale of the pseudopotential, while at the same time the property of exponential convergence is lost and replaced by algebraic convergence \cite{werner_general_2012,jeszenszki_accelerating_2018}. Extrapolation to the limit of zero range interactions has been successfully demonstrated for up to 66 fermions in three dimensions with an auxiliary-field quantum Monte Carlo approach \cite{Carlson2011}. 
Recently developed approaches like the transcorrelated method for short-range interactions 
can  further improve the power-law scaling \cite{jeszenszki_accelerating_2018}. 
Extending the applicability of this method to trapped systems is a promising avenue for future work.

\section{Acknowledgement}
We thank Tal Levy for initial work on this project, which indicated the inadequacy of conventional basis set expansions and motivated the analysis presented in this paper. This work was supported by the Marsden Fund of New Zealand
(Contract No.\ MAU1604), from government funding managed by the
Royal Society Te Ap\=arangi.

\appendix

\section{Fock-Darwin orbitals
%Single-particle basis: 
%Eigenfunctions of the harmonic oscillator and the angular momentum operator 
\label{app:basis}}

%  we scale down the length scale of the harmonic oscillator in order to have a better description for a small length scale of the interaction, but we also keep it large enough to describe the largest length scale of the system,
%\begin{eqnarray*}
%\frac{H^u_{\mathrm{osc}}}{\hbar \omega} &=& \sum_{\sigma}\sum_i^{N_\sigma}  \left(  -\frac{l_u^2}{2} \nabla^2_{i \sigma}  \ + \ \frac{m \omega_u^2}{2} r_{i \sigma}^2  \right) \ , 
%\end{eqnarray*}
%where $l_u$ is the modified length scale of the harmonic oscillator.
In the main text of the paper, we discussed the convergence properties from an analytic point of view, where  a product basis of one-dimensional basis functions provides an intuitive picture for the analysis. 
For numerical calculations it is more advantageous to apply a set of orbitals
that satisfy the symmetries of the system. This helps to restrict the problem to a single
irreducible representation of the symmetry operator, which reduces the required
number of the many-body basis functions and thus the requirements for computer memory and CPU time. 

We here use simultaneous eigenfunctions of the harmonic oscillator and the angular momentum operator known as Fock-Darwin orbitals
%As we mentioned earlier we used the eigenfunction of the harmonic oscillator, where we choose those basis functions, which are also eigenfunctions of the angular momentum operator,   
\begin{eqnarray*}
%H_{\mathrm{osc}} \, \varphi_{n,\ell} &=& \epsilon_{n \ell} \,  \varphi_{n \ell} \ , \\
L \, \varphi_{n \ell} &=&  \hbar \,  \ell \,  \varphi_{n \ell} \ , \label{angular}
\end{eqnarray*}
where $L$ is the angular momentum operator, %$\epsilon_{n \ell}$ is the single-particle energy, 
$\varphi_{n \ell}$ is the single-particle eigenfunction function, and $n$ and  $\ell$ are quantum numbers with non-negative integer and integer values. 
%The single-particle energy, $\epsilon^u_{n,\ell}$,  can be expressed with these quantum numbers explicitly,
%\begin{eqnarray}
%\epsilon^u_{n \ell} &=& \frac{\hbar \omega}{2} (2n+|\ell|+1) \ . \label{singlepartenerg}
%\end{eqnarray}
The eigenfunction $\varphi_{n \ell}$  can be
easily given in  polar coordinates,
\begin{align} \label{eq:FDorbital}
\nonumber \varphi_{n,\ell} (r, &\vartheta) = 
\sqrt{\frac{n! }{\tilde{l}^2 \pi (n+|\ell|)!}}  \left(\frac{r}{\tilde{l}}\right)^{|\ell|}  \\
& \hspace{2.5cm} \times  e^{-\frac{\left( r/\tilde{l} \right)^2}{2}}  e^{i \ell \vartheta}
{\mathcal L}_n^{|\ell|}\left(\frac{r^2}{\tilde{l}^2}\right) \ ,
\end{align}
 where  ${\mathcal L}_n^{|\ell|}(x)$ 
is the associated Laguerre polynomial. 
%By choosing parameter $\tilde{l}$ to be equal to the unit length of the harmonic oscillator $l=\sqrt{\frac{\hbar}{m \omega}}$,
%function $\varphi_{n,\ell}$ is the eigenfunction of $H_{\mathrm{osc}}$,

In the numerical calculation the finite single-particle basis set is chosen according to the total quantum number $\bar{n}=2n+\ell$, representing a ``shell.'' All  single-particle orbitals are selected where 
$\bar{n}$ is smaller than or equal to a maximal value $\bar{n}_\mathrm{max}$. 

The number $M$ of spatial single-particle orbitals can be expressed with $\bar{n}_\mathrm{max}$,
\begin{align}
M= \frac{\left(\bar{n}_\mathrm{max}+2\right) \left( \bar{n}_\mathrm{max}+1\right)}{2} \ .
\end{align}
In this paper the largest $\bar{n}_\mathrm{max}$ is 20, which 
corresponds to 231 spatial orbitals. The total number of the many-body basis functions for the three particles is equal to around
$6 \times 10^6$. Considering only those many-body states with projected angular momentum of $0\hbar$, %which 
%belongs to the ground state with angular momentum quantum number 1, 
the computational space can be reduced by about an order to $1.6 \times 10^5$ basis functions.

%the single-particle energy $\epsilon_{n \ell}$ in \rref{singlepartenerg}. Every single-particle state is included, where $\epsilon_{n \ell}$ is smaller or equal then a certain $\epsilon_{\mathrm{max}}$ value.

%Moreover the determinants are the eigenfunctions of the angular momentum operator, which decreases the number of the necessary basis functions by considering only a specific value of the angular momentum.

%\subsection{Scaled harmonic oscillator basis}

%In our case, we choose the single-particle functions 
%as an eigenfunction of the harmonic oscillator in two dimensions and the angular momentum operator ($L$),
%where $\epsilon_{n,\ell}$ is the energy, $n$ and $\ell$ are quantum numbers. The values of $n$ and $\ell$ are %non-negative integers and they explicitly provide the angular momentum [\rref{angular}] and $\epsilon_{n \ell}$,
%\begin{eqnarray}
%\epsilon_{n \ell} &=& \frac{\hbar \omega}{2} (2n+|\ell|+1) \ . \label{singlepartenerg}
%\end{eqnarray}
%Function $\varphi_{n \ell}$ is the single-particle function,
%\begin{eqnarray*}
%\varphi_{n,\ell} (r, \vartheta) &=& \sqrt{\frac{n! }{l^2 \pi (n+|\ell|)!}}  \left(\frac{r}{l}\right)^{|\ell|} %
%e^{-\frac{\left( r/l \right)^2}{2}}  e^{i \ell \vartheta}
%{\mathcal L}_n^{|\ell|}\left(\frac{r^2}{l^2}\right) \ ,
%\end{eqnarray*}
%where $l=\sqrt{\frac{\hbar}{m \omega}}$ is the unit length of the harmonic oscillator and ${\mathcal %L}_n^{|\ell|}(x)$ is the associated Laguerre polynomial.

\section{\label{appendixint}Evaluation of the  matrix elements}
 The matrix elements of the Hamiltonian can be evaluated according to the Slater-Condon rules \cite{slater_theory_1929,condon_theory_1930,slater_molecular_1931}, which express them as a linear combination of one-particle integrals
\begin{align*}
\langle \varphi_{n_1 \ell_1} | H_{\mathrm{osc}} | \varphi_{n_2 \ell_2} \rangle &= \int \mbox{d} {\bf r} \  \varphi_{n_1 \ell_1}^* ({\bf r}) \ H_{\mathrm{osc}} \  \varphi_{n_2 \ell_2} ({\bf r}) \ ,    
\end{align*}
and two-particle integrals
\begin{align*}
\langle \varphi_{n_1  \ell_1} \varphi_{n_2 \ell_2}| V | \varphi_{n_3  \ell_3} \varphi_{n_4  \ell_4} \rangle & \\
=\int \mbox{d} {\bf r}_1 \mbox{d}  {\bf r}_2 \, \varphi_{n_1  \ell_1}^* ({\bf r}_1) & \varphi_{n_2  \ell_2}^* 
({\bf r}_2) V({\bf r}_1-{\bf r}_2)  \varphi_{n_3  \ell_3}({\bf r}_1) \varphi_{n_4  \ell_4}({\bf r}_2) \ .
\end{align*}

The integrals are calculated with the single-particle basis described in Appendix \ref{app:basis}. The explicit expressions are given in the following sections. 

\subsection{Evaluation of one-particle integrals}

The one-particle integral can be evaluated analytically providing an easily implementable formula
\begin{align*}
 &\left \langle  \varphi_{n_1 \ell_1}  \left| \hat{H}_{\mathrm{osc}} \right|
\varphi_{n_2 \ell_2} \right \rangle  \\[0.5cm]
& \hspace{0.4cm} =\frac{\delta_{\ell_1 \ell_2} \hbar \omega}{2} 
{\Bigg (} \frac{1+\left(\tilde{l}/l\right)^4}{\left(\tilde{l} /l \right)^2}(2n_1+|\ell_1|+1) \delta_{n_1n_2}  \\
& \hspace{0.8cm} + \frac{1-(\tilde{l}/l)^4}{(\tilde{l}/l)^2} \sqrt{n_1(n_1+|m_1|)}\delta_{n_1,n_2+1} \\
& \hspace{0.8cm} + \frac{1-(\tilde{l}/l)^4}{(\tilde{l}/l)^2} 
\sqrt{(n_1+1)(n_1+|m_1|+1)}\delta_{n_1,n_2-1}  
 { \Bigg )} \ .
\end{align*}
 
% In a case of the unit length of the basis functions is equivalent to the unit length of the harmonic oscillator the matrix elements become diagonal. 

\subsection{Evaluation of two-particle integrals}

First, let us transform out the unit length of the harmonic oscillator  

\begin{align*}
\langle &\varphi_{n_1\ell_1}  \varphi_{n_2\ell_2} | -\frac{V_0 / \hbar \omega }{(R/l)^2 }
 e^{-\frac{|{\bf r}_2/l - {\bf r}_1/l|^2}{(R/l)^2}}
| \varphi_{n_3\ell_3}  \varphi_{n_4\ell_4}  \rangle  \\
& =\frac{l^2}{\tilde{l}^2}
\langle \varphi_{n_1\ell_1}  \varphi_{n_2\ell_2} |
-\frac{V_0 / \hbar \omega }{(R/\tilde{l})^2 }
 e^{-\frac{|{\bf r}_2/\tilde{l} - {\bf r}_1/\tilde{l}|^2}{(R/\tilde{l})^2}}
| \varphi_{n_3\ell_3}  \varphi_{n_4\ell_4}  \rangle \ ,
\end{align*}
which transfers the dependence of the unit length to a
scale factor. In the following we consider only the remaining matrix element on the right-hand side, where
both the basis function and  the operator have the same unit length.

The direct evaluation of the matrix elements with the single-particle orbitals $\varphi_{n\ell}$ defined in Appendix \ref{app:basis} is numerically unstable. Therefore, the 
integrals are calculated in Cartesian orbitals
\begin{eqnarray}
 \phi_{n_xn_y} (x,y)&=&   \chi_{n_x} (x) \chi_{n_y} (y) \ , \label{Cartesiandef} \\
 \chi_{n} (x) &=& \frac{1}{\sqrt{\sqrt{\pi}2^n n!\tilde{l}}} e^{-\frac{x^2}{2\tilde{l}^2}} \mathcal{H}_n(x/\tilde{l}) \ , 
\end{eqnarray}
where $\phi_{n_xn_y} (x,y)$ is the two-dimensional and $\chi_{n_x} (x)$ is the one-dimensional Cartesian function, and $\mathcal{H}_n(x)$ is the $n$th-order Hermite polynomial. The function $\phi_{n_xn_y} (x,y)$ can be used as a basis for expanding the single-particle basis
$\varphi_{n\ell}$ \cite{atakishiyev_finite_2001},
\begin{eqnarray}
| \varphi_{n\ell} \rangle &=& \sum_{n_xn_y}^{n_x+n_y=2n+|\ell|} d_{n_xn_y}^{n\ell} | \phi_{n_xn_y} \rangle \ , \label{Cartesiantrafo}
\end{eqnarray}
where $d_{n_x,n_y}^{n,\ell}$ is the Wigner small $d$ matrix \cite{landau_course_1977}. Then the two-particle integral in the basis
of $\varphi_{n \ell}$ can be calculated with multiple
unitary transformations
\begin{align}
\langle \varphi_{n \ell} &\varphi_{m \ell'}| V | \varphi_{p\ell''} \varphi_{q \ell'''} \rangle  \label{FDinteval}\\
& = \sum_{\substack{n_xn_ym_xm_y\\p_xp_yq_xq_y}}  {d_{n_xn_y}^{n\ell*}} {d_{m_xm_y}^{m\ell'*}} d_{p_xp_y}^{p\ell''} d_{q_xq_y}^{q\ell'''}
 \nonumber \\
& \hspace{2.5cm}  \times \langle \phi_{n_xn_y} \phi_{m_xm_y} | V | \phi_{p_xp_y} \phi_{q_xq_y} \rangle \ \nonumber , 
\end{align}
where the summation indices are restricted similarly to \rref{Cartesiantrafo}. Using \rref{Cartesiandef} the Cartesian integral can be separated according to the spatial variables
\begin{align}
\langle \phi_{n_xn_y} &\phi_{m_xm_y} | \hat{V} | \phi_{p_xp_y} \phi_{q_xq_y} \rangle \label{sepint} \\
&\hspace{2cm} =-\frac{V_0 \tilde{l}^4}{\pi R^2} I^{n_x m_x}_{p_x q_x} I^{n_y m_y}_{p_y q_y } \ , \nonumber
\end{align}
where the tensor $I^{nm}_{pq}$ can be given as
\begin{align*}
I^{nm}_{pq} &= \frac{1}{\sqrt{2^{n+m+p+q}(n!)(m!)(p!)(q!)}}   \\
& \hspace{0.2cm} \times \int\limits_{-\infty}^{\infty} \mbox{d} x_1  \int\limits_{-\infty}^{\infty} \mbox{d} x_2 \ 
\mathcal{H}_n(x_1) \mathcal{H}_m(x_2) \mathcal{H}_p(x_1) \mathcal{H}_q(x_2)  \cdot \\ 
& \hspace{4.5cm} \times e^{-\left(x_1^2+x_2^2\right)} e^{-\frac{\left(x_1-x_2\right)^2}{R'^2}} \ , 
\end{align*}
where $R'=R/l$. The integral above can be calculated analytically.

The recent work of Ref.~\cite{mujal_quantum_2017} used analytical expressions containing factorials, which can lead to numerical uncertainties even at small basis set sizes. Here we present a  different, iterative algorithm that can reach high accuracy using standard floating-point arithmetic.

Using the expansion of the Hermite polynomials
\begin{align*}
 \frac{\mathcal{H}_n(x)}{2^n n!} = \sum_i^n h_i^n x^i \ ,
\end{align*}
the tensor $I^{nm}_{pq}$ can be expressed with the following summations:
\begin{align}
I^{nm}_{pq} = & \delta_{ \mbox{\tiny mod}(n+m+p+q ,2), 0} \,  \label{Ieval}\\
& \hspace{1cm} \times  \sum\limits_i^n \sum\limits_k^p \sum\limits_j^m \sum\limits_l^q h_i^n h_k^p  h_j^m h_l^q
g_{i+k,j+l} \ , \nonumber  \\
g_{a,b}= & \int\limits_{-\infty}^{\infty} \mbox{d} x_1  \int\limits_{-\infty}^{\infty} \mbox{d} x_2 \, 
x_1^a x_2^b \, e^{-\left(x_1^2+x_2^2\right)} e^{-\frac{\left(x_1-x_2\right)^2}{R'^2}} \ . \label{calcgab}
\end{align}
Although integral \rrefsb{calcgab} can be evaluated analytically, the summation
in \rref{Ieval} contains the difference of large numbers, which decreases the 
numerical accuracy. In order to improve the numerical determination, we expand 
$g_{a,b}$ as a sum of $g_{r,0}$,
\begin{eqnarray}
g_{a,b} &=&  \sum_{r=a}^{a+b} e_r^{ab} g_{r,0}  \, \delta_{\mathrm{mod}(r,2),0}\ , \label{gabexpand}
\end{eqnarray}
where $\mathrm{mod}(x,y)$ is the modulo function \cite{eric_w._weisstein_mod_2018} and  the coefficients $e_r^{ab}$ can be obtained with a recursive algorithm:
\begin{eqnarray}
e_r^{a,b} &=& \frac{(b-1)R'^2 e_r^{a,b-2}+e_r^{a+1,b-1} }{2R'^2+1}  \ , \label{recgab} \\
e_a^{a,0} &=& 1 \ ,  \label{ea0} \\
e_a^{a-1,1} &=&\frac{1}{2R'^2+1} \ .  \label{ea1} 
\end{eqnarray}
Equations \rrefsb{recgab} - \rrefsb{ea1} can be derived with integration by parts from 
integral \rrefsb{calcgab}. 

Let us substitute \rref{gabexpand} into \rref{Ieval}
\begin{align}
I^{nm}_{pq} = & \delta_{ \mbox{\tiny mod}(n+m+p+q ,2), 0} \sum\limits_i^n \sum\limits_k^p \sum\limits_j^m \sum\limits_l^q h_i^n h_k^p  h_j^m h_l^q \, 
\nonumber \\
& \hspace{1.5cm} \times 
\sum_{r=i+k}^{i+k+j+l} e_r^{i+k,j+l} g_{r,0} \, \delta_{\mathrm{mod}(r,2),0} \ , \label{Ieval2}
\end{align}
where the obtained expression is still numerically unstable due to alternating signs of the 
coefficients $h_i^n$ and the increasing value of $g_{r,0}$ with $r$. In order to 
alleviate these numerical inaccuracies we extend the definition of the coefficient
$e_r^{i+k,j+l}$ to smaller indices of $r$:
\begin{eqnarray}
e_r^{i+k,j+l} = 0 \  \hspace{1cm}  \mbox{if} \hspace{1cm}   0 \le r < i+k \ . \label{esmallr}
\end{eqnarray}
Hence, the summation in \rref{Ieval2} can be reordered and all of the coefficients can 
be wrapped into the coefficient $D_r$:
\begin{align}
I^{nm}_{pq} = & \delta_{ \mbox{\tiny mod}(n+m+p+q ,2), 0} \, \sum_{r=0}^{n+m+p+q} D_r \, g_{r,0} \ , \label{Ieval3} \\
 D_r= & \sum\limits_i^n \sum\limits_k^p \sum\limits_j^m \sum\limits_l^q h_i^n h_k^p  h_j^m h_l^q
 e_r^{i+k,j+l} \ . \label{Ddef}
\end{align}
The relation between the neighboring $g_{a,0}$ can be determined with integration by parts of integral \rrefsb{calcgab}:
\begin{eqnarray}
g_{a,0} &=&\frac{(a-1)(2R'^2+1)}{4R'^2+4}g_{a-2,0} \ . \label{recga0} 
\end{eqnarray}
The numerical accuracy of the summation in \rref{Ieval3} can be increased if we apply 
relation \rrefsb{recga0} and evaluate 
the coefficient $D_r$ starting from the largest index:
%\todo{check revised wording}
\begin{align}
\tilde{D}_{r-2} &= \frac{(r-1)(2R'^2+1)}{4R'^2+4} \tilde{D}_r+ D_{r-2}, \label{Dtilde}\\
\tilde{D}_{n+m+p+q} &=  D_{n+m+p+q} \ . \label{Dtilde2} 
\end{align}
The coefficient $ \tilde{D}_{0}$ provides a simple expression for \rref{Ieval2},
\begin{align}
I^{nm}_{pq} = & \delta_{ \mbox{\tiny mod}(n+m+p+q ,2), 0} \, \tilde{D}_{0} \, g_{0,0}
\ , \label{Ieval4}
\end{align}
where $g_{0,0}$ can be determined by explicitly integrating integral \rrefsb{calcgab}:
\begin{eqnarray}
g_{0,0} &=& \frac{R'}{\sqrt{R'^2+1}} \label{g00} \ .
\end{eqnarray}

For determining the two-particle integrals on the computer, we use the following algorithm:
%Finally, let us conclude the determination of the 
%two-particle integrals in the computer algorithm. 
First, we determine the coefficients $e_r^{i+j+k+l}$ with 
\rrefsa{recgab}--\rrefsb{ea1} and \rref{esmallr}. After that, we calculate the coefficients 
$D_r$ and $\tilde{D}_r$ with \rrefsa{Ddef}, \rrefsb{Dtilde}, and \rrefsb{Dtilde2}. Then
the tensor $I^{nm}_{pq}$ can be determined from \rrefsa{Ieval4} and \rrefsb{g00}, with which the two-particle integrals can be easily evaluated from 
\rrefsa{sepint} and \rrefsb{FDinteval}. 
With the described algorithm the two-particle integrals are accurate at least for the 
first eight decimal digits using standard quadruple-precision (128-bit) floating-point arithmetic, where the maximal total quantum number ($2n+\ell$) is set to 20 for the single-particle basis function $\varphi_{n \ell}$.

\bibliography{2Drealization,book,scattering,Bose_gas,Fewfermions,Fewbosons,FCI,renormalization,Bethe_Peiers,Fermi_gases,Bertsch_parameter,DVR,FCIQMC,math,refs_jb,Quantum_Hall_effect,transcorrelated}

\end{document}